\documentclass{article} 
\usepackage[final]{nips_2016}
\usepackage{graphicx, times}
\usepackage{hyperref}
\usepackage{url}
\usepackage{subfig}
\usepackage{natbib}

\title{Opioid Atlas: Mapping Access to Pain Medication}

\author{
  Kris Sankaran\\
  Department of Statistisc \\
  Stanford University\\
  Stanford, CA 94305 \\
  \texttt{kriss1@stanford.edu} \\
   \And
  Suzanne Tamang \\
  Stanford Center for Biomedical Research \\
  Stanford University\\
  Stanford, CA 94305 \\
  \texttt{stamang@stanford.edu} \\
  \AND
  Ami S. Bhatt \\
  Departments of Medicine \& Genetics\\
  Stanford University\\
  Stanford, CA 94305 \\
  \texttt{asbhatt@stanford.edu} \\
}

\begin{document}

\maketitle

\begin{abstract}
Opiates are some of the most effective pain relief medications available for patients suffering from cancer and surgery-related pain. Despite the affordability and effectiveness of these medications, access to opiates is highly geographically variable.  Pain researchers have attributed geographic variation to various factors including the fear of opioid addiction, diversion of legal opiods to the underground market and pharmaceutical industry influences. However, the extent to which there is inequity in untreated cancer and surgery-related pain is unknown. To help opioid investigators study these questions, we designed a tool, the Opioid Atlas, for exploring data on legal opioid consumption, by country and time, collected by the International Narcotics Control Board. Our design borrows ideas from the data visualization and multivariate statistics communities, especially the principles of linking and dimensionality reduction. Our work is relevant to policymakers and pain researchers who wish to systematically assess country-level factors that contribute to differences in opioid access for patients with cancer and surgery-related pain.  The Opioid Atlas, and the code behind it, is freely available with an open source license.  
%
%
%

\end{abstract}

\section{Introduction}
Opiates are low cost and effective, an example of high-value healthcare.  Despite this fact, access to pain relief medications for patients suffering from cancer and surgery-related pain is limited and highly geographically variable \cite{knaul2015closing2, knaul2015closing}. Experts have suggested that international opiate policy reform is necessary to improve pain management and reduce unnecessary suffering, especially at the end-of-life \cite{berterame2016use}.  However, our scientific understanding of the factors that lead to disparities are limited to anecdotal evidence and healthcare surveys conducted on a relatively small sample of clinicians in a few countries \cite{laxmaiah2011effectiveness}. 

Our work aimed to achieve three goals.  The first was to obtain and systematically analyze a dataset describing trends in global opioid distribution.  The second was to develop a transformation of the dataset that would facilitate its use by other investigators.  Finally, we aimed to create a data visualization tool that a domain expert could use to explore patterns in global opioid access data across times, countries, and drugs.

In this paper, we describe our process, from data acquisition to tool production, and highlight several new insights into global opioid distribution that were enabled by the Opioid Atlas (https://krisrs1128.github.io/OpioidAtlas/).  Our R code for data processing, all analyses, and visualizations are freely available on github (https://github.com/krisrs1128/OpioidAtlas) under an open sources license.
%
%
%

\section{Methods}
\subsection{Data Sources}
Our global opioid distribution dataset was obtained from the International Narcotics Control Board (INCB).  The INCB is an independent and quasi-judicial monitoring body for the implementation of the United Nations international drug control conventions. It was established in 1968 in accordance with the Single Convention on Narcotic Drugs, 1961 \cite{who1961}. Every year, they publish technical reports on the state of opioid consumption, including tables of raw data.

This data is spatiotemporal, with one sample for each country for every year from 1989 to 2013, with an extra dimension for drug type. We note a few nonstandard characteristics.  First, there are many zeros, mostly associated with smaller countries with no access to certain drugs or incomplete reporting, which tend to occur earlier in the collection period and over consecutive years.  Second, there is a dramatic difference in scale between the largest and smallest consumers -- plotting all series on the same y-axis range is not informative.  Third, the different classes of drugs are not directly comparable, because some have stronger medical potency than others. Hence, we transformed each series into "morphine-equivalent" kilograms per year \cite{cdcequivalents}.  Finally, there are many departures from smoothness, across time, space, and drugs, in the sense that dramatic dips and spikes occur within individual countries time series, scales can shift substantially between neighboring countries, and countries can have variable access to the different opioids.

\subsection{Exploratory Techniques}
The Opioid Atlas is designed to give representations of the data that facilitate comparisons across times, countries, and drugs. Our approach is to simplify the navigation across a large collection of related time series. To this end, we borrow ideas from the data visualization and multivariate statistics communities, especially the principles of linking and dimensionality reduction.

\subsubsection{Linking}
In the visualization community, linking refers to tying together different representations of a dataset in one display \cite{buja1996interactive, becker1987brushing}. Our first view links a choropleth of opioid access with time series of opioid consumption for individual countries. The linking is implemented through mouseovers: when the user hovers over a country, the time series display transitions to reflect abundance across all drugs for the hovered-over country. When the user selects a year and an opioid, the choropleth and time series update to reflect the new specification. This linking makes it easy to compare time series across neighboring countries.

In the next few views, we generalize this linking idea by defining our own similarities across countries. This makes it easier to compare countries that have similarly shaped series but which are not necessarily geographically close. 

\subsubsection{Cognostics}
In the first generalization, we adapt the idea of cognostics, an early exploratory data analysis technique for comparing large collections of scatterplots \cite{tukey1977exploratory, hafen2013trelliscope}. A cognostic is a single summary statistic for a scatterplot -- correlation or regression slope, for example -- which can be used to order the otherwise unwieldy collection. For the Opioid Atlas, we precomputed a collection of cognostics -- the difference between the last and first measured point and the maximum one year increase, for example -- for each country and each drug. We then generate a dot-plot of these cognostics and link each dot to the time series view for that country. This allows the user to see both typical and outlier values for each cognostic, as well as the associated original series.

\subsubsection{Multidimensional Scaling}
In the second generalization, we use multidimensional scaling (MDS), a method from multivariate statistics, to organize the individual time series. In multidimensional scaling, we define an abstract distance between samples and generate a scatterplot where the distances between points approximate the original abstract distance \cite{kruskal1964multidimensional}. In the Opioid Atlas, our abstract distance is the Euclidean distance between cube-rooted time series. The resulting MDS scatterplot provides an alternative to the world map in the first view. As in this first view, we link individual points to their associated time series. This facilitates comparison between series that are shaped similarly, according to raw Euclidean distance.

\subsubsection{Local Regressions}
In our final generalization, we define a new map according to overall and trends in access to opioids. The goal is to define an MDS-like scatterplot directly in terms of quantities that are interesting to domain experts. Towards this, we define a "local" average and slope for each series and each time point. For each fixed year, we display a scatterplot of countries and drugs with respect to these statistics. As before, each point is linked with the associated time series view. Changing years transitions points on the scatterplot to reflect updated statistics. We define local averages and slopes using local ridge regressions \cite{cleveland1988locally}, calling fitted values at each timepoint local averages, and the associated slopes the local slopes. In this view, it is easy to see groups of countries with rapid increases and decreases in access across short time periods, while maintaining context about overall access to opioids.

\section{Results}
Here we highlight two illustrative examples of how we used the Opioid Atlas to identify interesting trends in global opioid distribution.  Our examples highlight the types of relationship that are of interest to opioid investigators and would not be revealed by traditional opioid distribution analysis and reporting methods.  

\begin{figure}[ht]%
    \centering
    \subfloat[Sovereignity]{{\includegraphics[width=3.75cm]{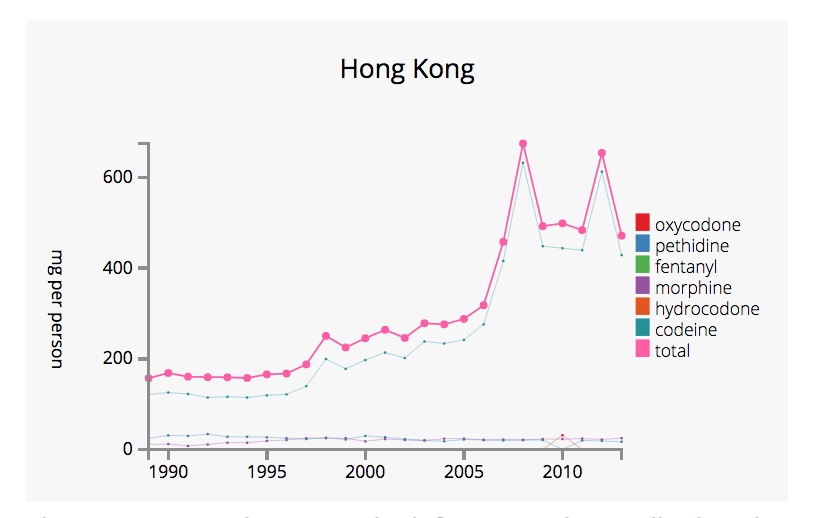}}}%
    \qquad
    \subfloat[Drug Toggling]{{\includegraphics[width=6.5cm]{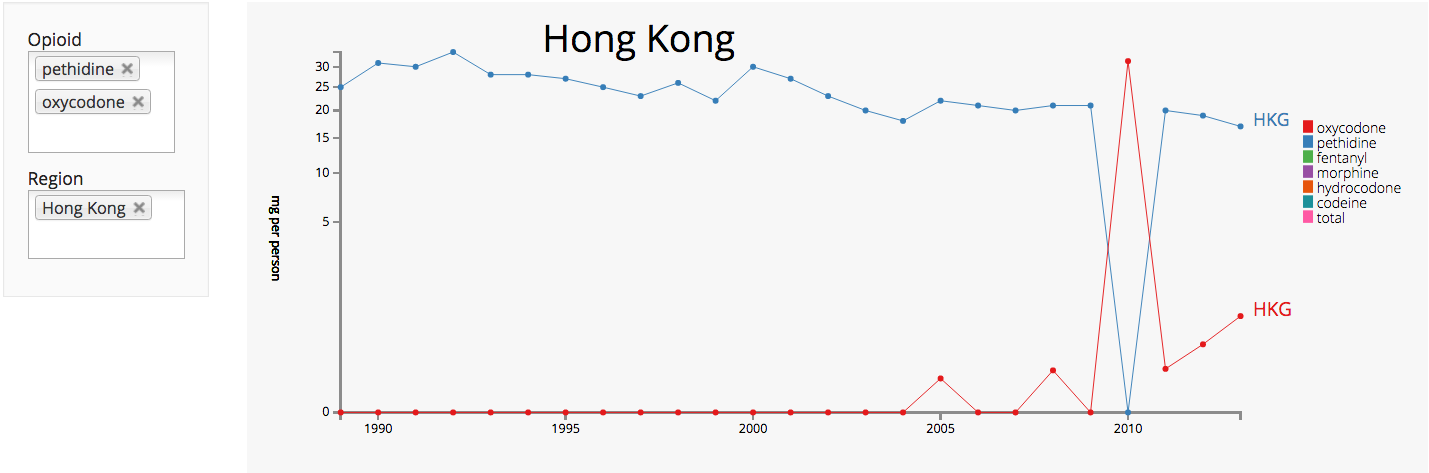} }}%
    \caption{Temporal opioid consumption trends in Hong Kong. Annual consumption trends before and after the transfer of sovereignty from the UK to China (1997) appear on the left; an example of toggling between oxycodone and pethidine in 2010 appears on the right.}%
    \label{fig:examplehk}%
\end{figure}

We used our first generalization to prioritize our search for temporal trends of interest at the country level.  One trend was discovered by identifying Asian countries with rapidly increasing levels of oxycodone consumption and reviewing their linked time series.  Figure \ref{fig:examplehk} shows the linked time series for Hong Kong, which shows dramatic increases in total annual opioid consumption after the transfer of sovereignty from the United Kingdom to the People's Republic of China in 1997. Also illustrated in Hong Kong's time series, was evidence of "toggling."  Opioid class toggling is characterized by a decrease in one opioid class due to an increased consumption of opioid class, within the same duration.  An example demonstrating opioid class toggling between oxycodone and pethidine is shown on the right of Figure \ref{fig:examplehk}.

\begin{figure}[ht]
    \centering
  \includegraphics[width=10cm]{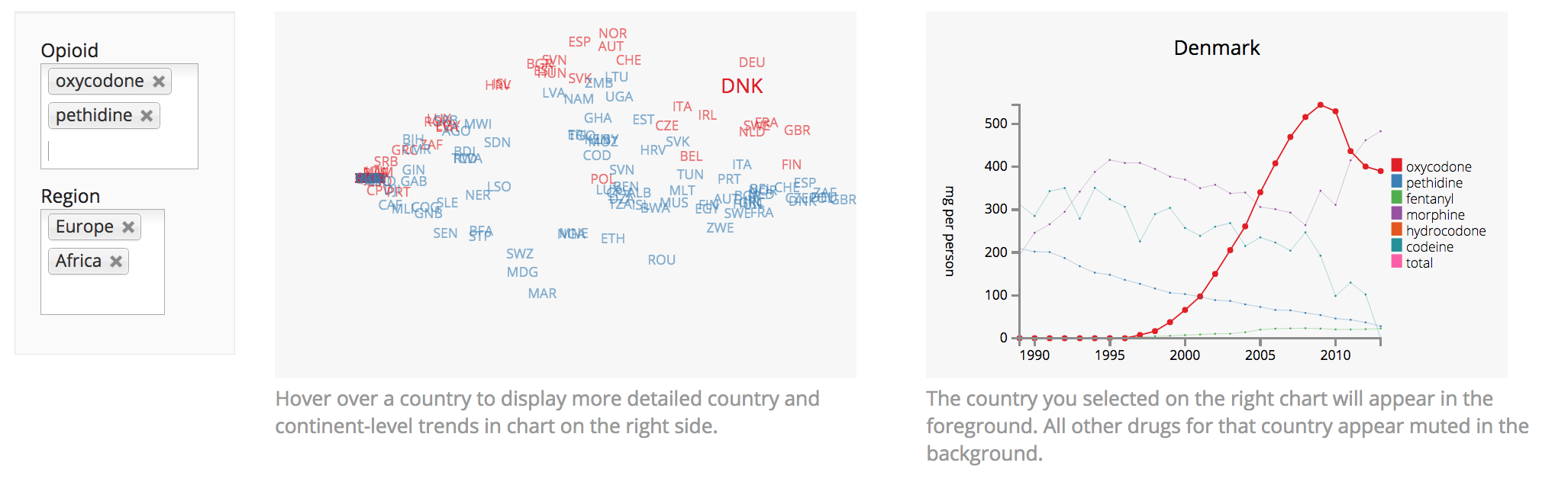}
  \caption{Opioid consumption clusters in Europe and Africa with linked Danish time series on the right.}
  \label{fig:dk}
\end{figure}

For two opioid classes, oxycodone and pethidine, we used our second generalization, based on MDS, to automatically group European and African countries with similar consumption patterns.  Figure \ref{fig:dk} shows the  country-level ``clusters" we found for the opioid oxycodone, in red, and the opioid pethidine, in blue.  Our linked country-level time series revealed that red countries located in the upper right side of the chart, (e.g., Denmark, Germany, Ireland), were associated with increased oxycodone consumption, and exclusively European.  The linked time series of countries that in the upper middle of the chart in blue (e.g., Zambia, Uganda, Namibia, Ghana, and Lithuania) showed increased pethidine consumption; although they include some European countries such as Lithuania and Latvia, most are African. 

Lastly, for the European countries associated with rapid increase in oxocodone consumption in Figure \ref{fig:dk}, there are two specific findings that may warrant consideration by opioid investigators.  Specifically, most European countries with a rapid increase in oxycodone consumption showed a rapid decrease for the opioids pethidine, morphine and hydrocodone (i.e., toggling).  Notably, Denmark was the only country with a visual decrease in oxycodone consumption shortly after 2007, when three of the drug manufacturer's executives pleaded guilty in federal court to criminal charges that they misled regulators, doctors and patients about the drug’s risk of addiction and its potential to be abused \cite{meier2007guilty}.

\section{Discussion}
In order to identify the best strategies to improve pain management for cancer and surgery patients, a better understanding of country and regional consumption patterns, pharmaceutical industry influences, and sociopolitical factors that impact country-level consumption are needed.  The Opioid Atlas was developed to facilitate description of historical patterns.  Based on the INCB's global opioid distribution data from 1989-2013, our tool allows for comparisons across time, regions, countries, and opioid classes. 

Our results show how the atlas can be used to identify relationships of interest to opioid investigators.  We highlight trends in country-level opiate consumption that are not fully explained by population size, gross national product or healthcare access more generally.  For example, the dramatic increase in opioids observed in the five year period after Hong Kong's sovereignty was transferred from the United Kingdom to the People's Republic of China, an observation of interest to investigators who have hypothesized an association between low opioid use and British colonization.  We also highlight and example of drug-class toggling -- i.e., in the absence of new randomized clinical trial evidence or a regulatory recall, the replacement of one class for another could suggest possible pharmaceutical industry influences, which would not be seen by analyzing opiate consumption in aggregate. 

It is important to note that our tool is only the first step of a scientific inquiry process that must involve further studies to provide a more certain explanation for opioid disparities.  Additional limitations of our work include the quality of the INCB data, which can differ by country and reporting year.

\section{Conclusion}

We introduced the Opioid Atlas, a website of visualizations that can be used to better describe the history of opioid consumption globally, and described the process and principles behind its design. Our work is a case study in blending ideas from data visualization and multivariate statistics in order to build a tool adapted to a specific domain problem. Our focus on linking and interactivity comes from the visualization literature, while our application of dimensionality reduction methods has a distinctly statistical flavor.

We have discussed patterns in the consumption of opioids both to demonstrate the the ease of navigation enabled by this tool and to suggest its usefulness in practice. By making our data and code publicly available, we attempt to reflect best practices in open, reproducible research. By making this data more easily explorable, we hope to inspire a data-centered discussion, both among domain experts and across the general public, of disparities in access to pain relief.

\subsubsection*{Acknowledgments}
We would like to thank Stefano Berterame from the INCB for providing the global opioid distribution dataset and encouraging the development of the Opioid Atlas.

\bibliographystyle{plainnat}
\bibliography{opioid_atlas}

\end{document}